\documentclass[12pt]{article}
\pdfoutput=1
\usepackage{subfigure}
\usepackage{amssymb,amsmath}
\usepackage{graphicx}
\usepackage{color}
\usepackage[colorlinks=true
,urlcolor=blue
,citecolor=blue
,linkcolor=blue
,pagecolor=blue
,linktocpage=true
,pdfproducer=medialab
]{hyperref}
\usepackage[a4paper,width=15.2cm]{geometry}
\makeatletter \renewcommand{\@dotsep}{10000} \makeatother

\newcommand{\beq}{\begin{equation}}
\newcommand{\eeq}{\end{equation}}
\newcommand{\bea}{\begin{eqnarray}}
\newcommand{\eea}{\end{eqnarray}}

\begin{document}

\begin{center}

 {\Large\bf   Revisiting mGMSB in light of a 125 GeV Higgs
 } \vspace{1cm}

{\large   M. Adeel Ajaib\footnote{ E-mail: adeel@udel.edu}, Ilia Gogoladze\footnote{E-mail: ilia@bartol.udel.edu\\
\hspace*{0.5cm} On  leave of absence from: Andronikashvili Institute
of Physics, 0177 Tbilisi, Georgia.},    \vspace{.3cm}Fariha Nasir\footnote {E-mail: fariha@udel.edu }  and  Qaisar Shafi\footnote{ E-mail:
shafi@bartol.udel.edu} } \vspace{.9cm}

{\baselineskip 20pt \it
Bartol Research Institute, Department of Physics and Astronomy, \\
University of Delaware, Newark, DE 19716, USA  } \vspace{.5cm}

\vspace{1.5cm}
 {\bf Abstract}
\end{center}

We explore the implications of a $124-126$ GeV CP-even Higgs boson on the fundamental parameter space and sparticle spectroscopy of the minimal gauge mediated supersymmetry breaking (mGMSB) scenario. The above mass for the Higgs boson yields stringent lower bounds on the sparticle masses in this class of models. The lightest neutralino and stau masses lie close to 1.5 TeV and 800 GeV respectively, while the majority of the sparticle masses are in the several to multi-TeV range. We show that with a single pair of $5+\overline{5}$   SU(5) messenger multiplets, the lower limit on the gravitino mass is  $\sim 360  {\rm \ eV}$. This is reduced to about $60  {\rm \ eV}$ if five pairs of $5+\overline{5}$   messenger fields are introduced. Non-standard cosmology and non-standard gravitino production mechanisms are required in order to satisfy cosmological observations.

\newpage

\renewcommand{\thefootnote}{\arabic{footnote}}
\setcounter{footnote}{0}



\section{Introduction  \label{intro}}

Recently, some evidence for a SM-like Higgs boson with mass $ \sim 125$ GeV has been reported by
the ATLAS and CMS experiments \cite{atlas_h,cms_h}.  The results presented at the Moriond 2012 conference for a combined Tevatron analysis with  $10^{-1}$ fb integrated luminosity also support the LHC excess corresponding to a Higgs mass of around 125 GeV \cite{Moriond}. A Higgs boson with  $m_h \sim 125$ GeV
 places stringent constraints on supersymmetry (SUSY), especially in the context of  the minimal  supersymmetric standard model (MSSM)
\cite{Gogoladze:2011aa,Baer:2011ab, Arbey:2011ab,Draper:2011aa}.  In order to realize a light CP-even Higgs of mass around 125 GeV in the MSSM, we need either a very large, $O (10-100)$ TeV,  stop quark mass, or a large trilinear soft supersymmetry  breaking (SSB) A-term with stop quark mass still around a TeV  \cite{Gogoladze:2009bd}.  Assuming gravity mediated SUSY breaking, it was shown in   ref. \cite{Gogoladze:2011aa} that a SM-like Higgs boson with mass
$\sim 125  {\rm \ GeV}$ is nicely accommodated in SUSY  grand unified theory (GUT) models with $t$-$b$-$\tau$ Yukawa coupling unification at $M_{\rm GUT}$ \cite{big-422}.

Models with gauge mediated SUSY breaking (GMSB) model provide a compelling resolution of the  SUSY flavor problem, as a consequence of the flavor blind gauge interactions responsible for generating the SSB term \cite{Giudice:1998bp}.    In both the minimal \cite{Giudice:1998bp} and general \cite{Meade:2008wd}  GMSB scenarios  the trilinear SSB A-terms are relatively small  at the messenger scale, even if an additional sector is added to generate the $\mu/B\mu$ terms \cite{Komargodski:2008ax}. Although non-zero A-terms are generated at the low scale through renormalization
group equation (RGE) running, a significantly high scale for the messenger fields or very heavy gauginos are required, thereby making most of  the sparticles very heavy and difficult to access at the LHC.

In this paper we  revisit the minimal GMSB (mGMSB)  model in light of a SM like Higgs with  mass around 125 GeV.  Several studies \cite{Arbey:2011ab, Draper:2011aa} analyzing the GMSB scenario have recently appeared. In this paper we   perform a more  comprehensive study of the mGMSB model
 by scanning all the essential  parameters  characteristic of this scenario.
  The messenger scale is allowed to be as high as  $10^{16}$ GeV and we study  the resulting sparticle spectrum
  corresponding to the SM-like  light CP-even Higgs mass of  $m_h=125 \pm 1$ GeV.

The layout of this  paper is as follows. In Section \ref{model} we briefly summarize  the mGMSB model and the relevant soft supersymmetry breaking terms.
Section \ref{constraintsSection}  summarizes the scanning procedure and the
experimental constraints we employ. In Section \ref{results} we present our results, focusing in particular on the
 sparticle mass spectrum. The  table in this section presents some benchmark points which summarize  the prospects of testing these predictions  at the LHC. Our conclusions are presented in Section \ref{conclusions}.


\section{Minimal GMSB  and SSB Terms\label{model}}

Supersymmetry breaking in a typical GMSB scenario takes place in a hidden sector, and this effect is communicated to the visible sector via messenger fields. The messenger fields interact with the visible sector via known SM gauge interactions, and induce the SSB terms in the MSSM through loops.
In order to preserve perturbative gauge coupling unification, the minimal GMSB   scenario  can include $n_5$  $5_i+\overline{5}_i$  ($i=1, ... , n_5$)  or a single $10+\overline{10}$ \cite{Giudice:1998bp}, or $10+\overline{10}+5 +\overline{5}$, or  $15+\overline{15}$  \cite{Joaquim:2006uz} multiplets of  SU(5).  For simplicity, we only consider  the case with $n_5$ $5_i+\overline 5_i$ vectorlike multiplets.  Notice that $5+\overline 5$  includes $SU(2)_L$ doublets $(\ell +\bar{\ell})$, and  $SU(3)_c$ triplets $(q+\bar q)$.
In order to incorporate SUSY breaking in the messenger sector,  the fields in ($5+\overline 5$) multiplets are coupled, say, with the hidden sector gauge singlet chiral field $S$:
\begin{equation}
W \supseteq y_1 S\, \ell \, \bar{\ell}+y_2 S\,  q \, \bar{q},
\end{equation}
where $W$ denotes the appropriate superpotential. Assuming non-zero vacuum expectation values (VEVs) for the scalar and $F$ components of $S$, namely $S=\langle S \rangle+\theta^2 \langle F \rangle$, the mass spectrum of the messenger fields is as follows:
\begin{equation}
  m_b = M \sqrt{ 1 \pm {\Lambda \over M} }, \ \
m_f = M.
\end{equation}
  Here $m_b$ and $m_f$ denote the masses of the bosonic and fermionic components  of the  appropriate messenger superfield, $M = y \langle S \rangle$ and
  $\Lambda = \langle F \rangle / \langle S \rangle$.
The dimensionless parameter $\Lambda /M$ determines the  mass splitting between the scalars and fermions in the messenger multiplets.
This breaking is transmitted to the MSSM particles via loop corrections.

 At the messenger scale the MSSM gaugino masses are generated at  1-loop level, and assuming $\langle F \rangle \ll \langle S \rangle^2$,  are given by
\begin{equation}
M_{i} = n_5\, \Lambda \, \frac{\alpha_i}{4\pi}  ,
\label{gauginomass}
\end{equation}
 where $i=1,2,3$  stand for the  $SU(3)_c$,  $SU(2)_L$  and  $U(1)_Y$,  sectors, respectively.
The MSSM scalar masses  are induced at  two loop level:
\begin{equation}
m^2(M) = 2\,  n_5\,  \Lambda^2 \,  \sum_{i=1}^3 \, C_i
  \left( \alpha_i \over 4 \pi \right)^2,
\label{scalarmass}
\end{equation}
where $C_1 = 4/3$,   $C_2 = 3/4$ and
$C_3 = (3/5) (Y/2)^2$, and $Y$ denotes the hypercharge.

 The A-terms in mGMSB models vanish  at the messenger scale (except when the MSSM and messenger fields mix \cite{Evans:2011bea}, which we do not consider in this study). They
 are generated from the RGE running below the messenger scale.

The mGMSB spectrum is therefore completely specified by the following parameters defined at the messenger scale:
\begin{equation}
M_{\mathrm{mess}}, \Lambda,  \mathrm{tan}\beta,  sign({\mu}),  n_5,  c_{\rm grav}.
\label{mgmsb-params}
\end{equation}
$M_{\mathrm{mess}}\equiv M$  and $\Lambda$ are the messenger and SSB mass scale defined above , and ${\rm tan\beta}$ is the
ratio of the VEVs of the two MSSM Higgs
doublets. The magnitude of $\mu$, but not its sign, is determined by the
radiative electroweak breaking (REWSB) condition. The parameter $c_{\rm grav} \geq 1$ effects the mass of the gravitino and we set it equal to unity from now on.


\section{Phenomenological Constraints and Scanning Procedure\label{constraintsSection}}

We employ the ISAJET~7.82 package~\cite{ISAJET}  to perform random
scans over the fundamental parameter space.
In this package, the weak scale values of gauge and third generation Yukawa
couplings are evolved to $M_{\rm mess}$ via the MSSM renormalization
group equations (RGEs) in the $\overline{DR}$ regularization scheme.
The various boundary conditions are imposed at
$M_{\rm mess}$ and all the SSB
parameters, along with the gauge and Yukawa couplings, are evolved
back to the weak scale $M_{\rm Z}$.
In the evaluation of Yukawa couplings the SUSY threshold
corrections~\cite{Pierce:1996zz} are taken into account at the
common scale $M_{\rm SUSY} = \sqrt{m_{{\tilde t}_L}m_{{\tilde t}_R}}$,
where $\tilde t_L$ and $\tilde t_R$
are the third generation left and right handed stop quarks.
 The entire
parameter set is iteratively run between $M_{\rm Z}$ and $M_{\rm
mess}$, using the full 2-loop RGEs, until a stable solution is
obtained. To better account for leading-log corrections, one-loop
step-beta functions are adopted for gauge and Yukawa couplings, and
the SSB parameters  are extracted from RGEs at multiple scales
$m_i=m_i(m_i)$. The RGE-improved 1-loop effective potential is
minimized at $M_{\rm SUSY}$, which effectively
accounts for the leading 2-loop corrections. Full 1-loop radiative
corrections are incorporated for all sparticle masses.

An approximate error of  $\pm 3$ GeV in the ISAJET estimation of the Higgs mass
largely arises from theoretical  uncertainties \cite{Degrassi:2002fi}  in the calculation   and
to a lesser extent from experimental uncertainties.

We perform random scans for the following range of the mGMSB parameter space:
\begin{eqnarray}
0\leq & \Lambda  &\leq  10^7 \ \, \rm{GeV} \nonumber \\
1.01 \Lambda \leq   & M_{\rm mess} & \leq  10^{16} \ \, {\rm GeV} \ \nonumber \\
1.5\leq &\tan\beta &\leq 60 \nonumber \\
 &\mu & > 0,
 \label{parameterRange}
\end{eqnarray}
with    $m_t = 173.3\, {\rm GeV}$  \cite{:1900yx}. We have checked that our results are not
too sensitive to one or two sigma variation in the value of $m_t$  \cite{Gogoladze:2011db}.
We use $m_b(m_Z)=2.83$ GeV which is hard-coded into ISAJET.

In performing the random scan a uniform and logarithmic distribution of random points is first generated in the parameter space given in Eq. (\ref{parameterRange}).
The function RNORMX \cite{Leva} is then employed
to generate a gaussian distribution around each point in the parameter space.  The points with CP-even Higgs mass in the range 125 $\pm$ 1 GeV are scanned more rigorously using this function.

 The data points collected all satisfy
the requirement of REWSB. After collecting the data, we impose
the mass bounds on all the particles \cite{Nakamura:2010zzi} and use the
IsaTools package~\cite{Baer:2002fv}
to implement the various phenomenological constraints. We successively apply the following experimental constraints on the data that
we acquire from ISAJET:
\begin{table}[h]\centering
\begin{tabular}{rlc}
$m_h~{\rm (lightest~Higgs~mass)} $&$ \geq\, 114.4~{\rm GeV}$          &  \cite{Schael:2006cr} \\
$BR(B_s \rightarrow \mu^+ \mu^-) $&$ <\, 4.5 \times 10^{-9}$        &   \cite{:2007kv}      \\
$2.85 \times 10^{-4} \leq BR(b \rightarrow s \gamma) $&$ \leq\, 4.24 \times 10^{-4} \;
 (2\sigma)$ &   \cite{Barberio:2008fa}  \\
$0.15 \leq \frac{BR(B_u\rightarrow
\tau \nu_{\tau})_{\rm MSSM}}{BR(B_u\rightarrow \tau \nu_{\tau})_{\rm SM}}$&$ \leq\, 2.41 \;
(3\sigma)$ &   \cite{Barberio:2008fa}  \\
 $ 0 \leq \Delta(g-2)_{\mu}/2 $ & $ \leq 55.6 \times 10^{-10} $ & \cite{Bennett:2006fi}
\end{tabular}\label{table}
\end{table}

Note that for $\Delta(g-2)_{\mu}$, we only require that the  model does no worse than the SM.


\begin{figure}[t!]
\begin{center}
\includegraphics[width=7.2cm,height=6.cm]{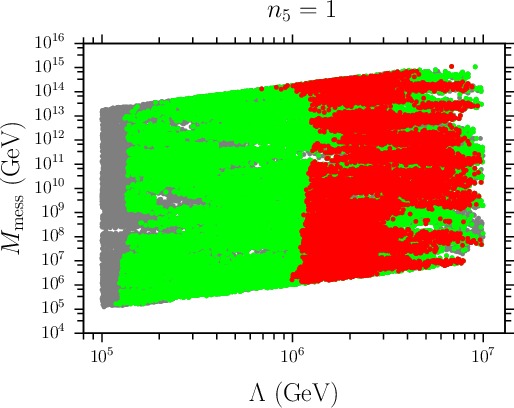}
\includegraphics[width=7.2cm,height=6.cm]{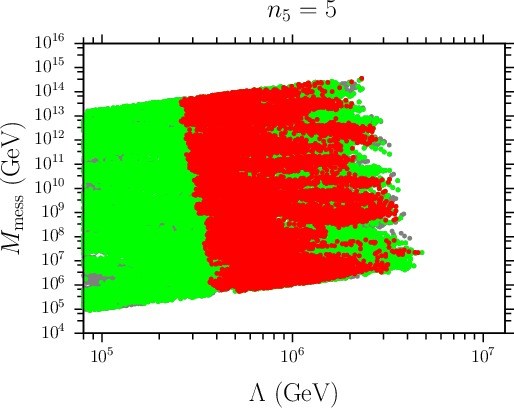}\vspace*{3mm}
\includegraphics[width=7.2cm,height=5.6cm]{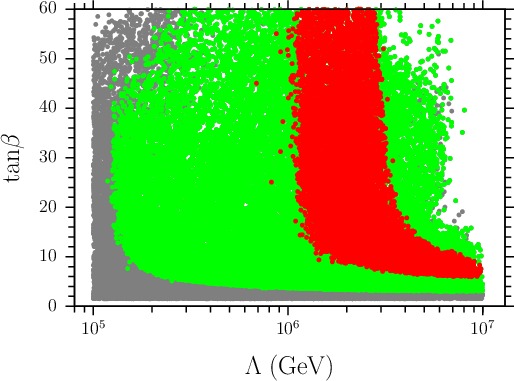}
\includegraphics[width=7.2cm,height=5.6cm]{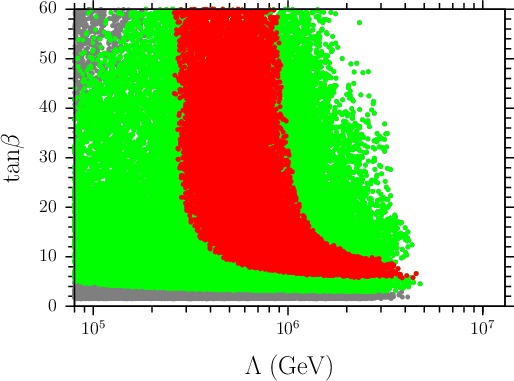}
\end{center}
\caption{Plots in $M_{\mathrm{mess}} - \Lambda$ and $ \mathrm{tan}\beta - \Lambda $ planes for  $n_5=1$ and $n_5=5$.
 Gray points are consistent with REWSB. Green points satisfy particle mass bounds and constraints described in section \ref{constraintsSection}.
  In addition, we require that green points do no worse than the SM in terms of $ (g-2)_{\mu} $.
  Red points belong to a subset of green points and satisfy the Higgs mass range 124 GeV $\leq m_h \leq 126$ GeV. }
\label{fundp}
\end{figure}


\section{Sparticle Spectroscopy \label{results}}

In this section, we present the sparticle spectroscopy which results from the procedure outlined in section \ref{constraintsSection}. We focus on the following mass range for the lightest CP-even SUSY Higgs boson:
\begin{eqnarray}
124 \mathrm{\ GeV} \lesssim m_h \lesssim 126 \mathrm{\ GeV}.
\label{mhrange}
\end{eqnarray}

In Figure \ref{fundp}, we show our results in the $M_{\mathrm{mess}} - \Lambda$  and $M_{\mathrm{mess}} - \mathrm{tan}\beta$ planes for $n_5=1$ and $n_5=5$.
The reason we choose these values for the number of messenger fields is that the sparticle spectrum, along with some other salient features, do not change appreciably for the intermediate values $n_5=2,\ 3$ and  4.  In the figures the gray points  are consistent with REWSB, whereas the green points, a subset of the gray ones, satisfy all the constraints described in section \ref{constraintsSection}. The red points correspond to the Higgs mass range given in equation (\ref{mhrange}), and form a subset of the green ones. We see that the desired Higgs mass requires relatively large values of $\Lambda$ which, in turn, pushes the MSSM sparticle mass spectrum to larger values. The minimum values of the parameters $M_{\rm mess}$ and $\Lambda$, for $n_5=1$,  and satisfying the bound  in Eq. (\ref{mhrange}) is $ \sim 10^6$ GeV. For $n_5=5$, the minimum values of  $M_{\rm mess}$ and $\Lambda$ are $ 5 \times 10^5 \rm \ GeV$ and $ 2.5 \times 10^5 \rm \ GeV$, respectively. These can have interesting effects on the  mass of the gravitino in mGMSB as we will see below.

  The $\mathrm{tan}\beta - \Lambda$ plot in Figure \ref{fundp} shows a trend in which the red points appear to merge as we  approach smaller values of $\mathrm{tan}\beta$ and large values of $\Lambda$, with a minimum value of $\mathrm{tan}\beta \sim 6$.  From this, we expect the  MSSM spectrum to be much heavier for smaller values of $\mathrm{tan}\beta$.  Note that varying $n_5$ from 1 to 5 does not change the range of $\mathrm{tan}\beta$  (red points) by much, while the values for
  $M_{\mathrm{mess}}$  and  $\Lambda$ are more significantly altered.

\begin{figure}[t!]
\begin{center}
\includegraphics[width=7.2cm,height=6.cm]{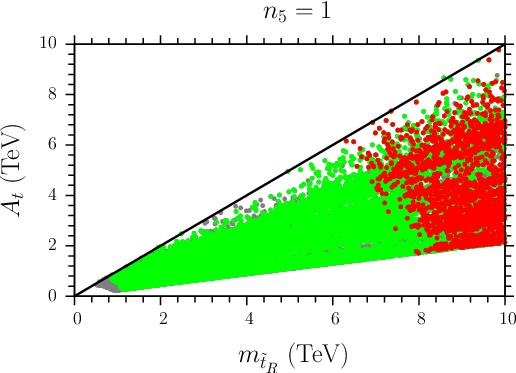}
\includegraphics[width=7.2cm,height=6.cm]{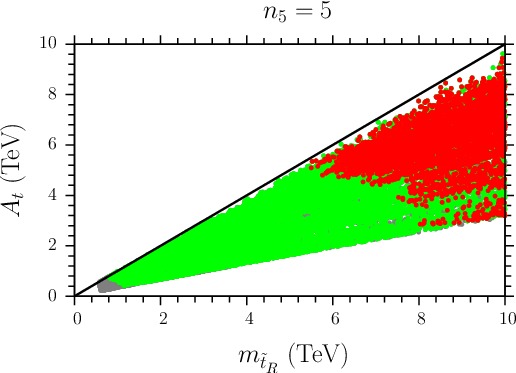}\vspace*{3mm}
\includegraphics[width=7.2cm,height=5.6cm]{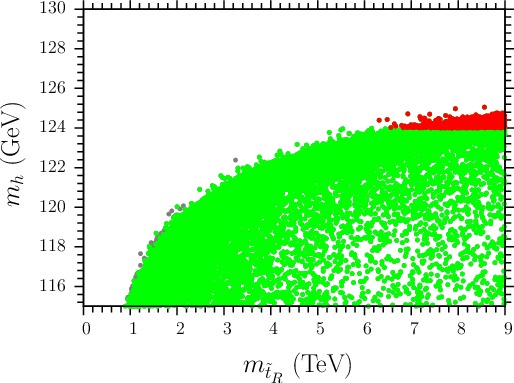}
\includegraphics[width=7.2cm,height=5.6cm]{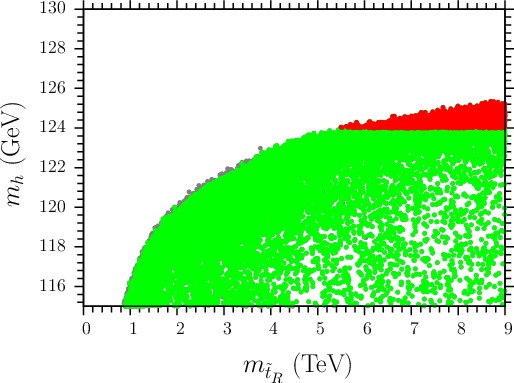}
\end{center}
\caption{Plots in $A_t - m_{\tilde{t}_R}$ and $m_h - m_{\tilde{t}_R}$ planes for $n_5=1$ and $n_5=5$. Color coding is the same as described in Figure \ref{fundp}. }
 \label{atmstop}
\end{figure}

We present in
Figure \ref{atmstop}  the results in $m_h - m_{\tilde{t}_R}$   and $A_t - m_{\tilde{t}_R}$ planes for $n_5=1$ and $n_5=5$. The color coding is the same as in Figure  \ref{fundp}. Note that since in the mGMSB scenario the non-diagonal elements in the squark and slepton mass matrices are  always smaller in comparison to the diagonal elements, we will use left and right handed notations for the third generation squark and slepton masses in our discussion.
The   $m_h - m_{\tilde{t}_R}$ panel indicates that the minimal value of  $m_{\tilde{t}_R}$, which corresponds to $m_h=125\pm 1$ GeV
 (red points) is above 6 TeV, for $n_5=1$.  It is more than $5$ TeV for $n_5=5$. The lower bound on $m_{\tilde{t}_R}$, therefore, is very large.

  In order to understand our finding, consider the one loop
contributions to the CP-even Higgs boson mass \cite{at}:
\begin{eqnarray}
\left[ m_{h}^{2}\right] _{MSSM} \approx M_{Z}^{2}\cos ^{2}2\beta \left( 1-\frac{3
}{8\pi ^{2}}\frac{m_{t}^{2}}{v^{2}}t\right)
+\frac{3}{4\pi ^{2}}\frac{m_{t}^{4}}{v^{2}}\left[ t+\frac{1}{2}X_{t} \right],  \label{mh}
\end{eqnarray}
where
\begin{eqnarray}
v=174.1 {\rm \ GeV}, \ \
t =\log \left(
\frac{M_{S}^{2}}{M_{t}^{2}}\right),~~~
X_{t} &=&\frac{2\widetilde{A}_{t}^{2}}{M_{S}^{2}}\left( 1-\frac{\widetilde{A}%
_{t}^{2}}{12M_{S}^{2}}\right). \label{A1}
\end{eqnarray}%
Also $\widetilde{A}_{t}=A_{t}-\mu \cot \beta $, where
$A_{t}$ denotes the stop left and stop right soft
mixing parameter and $M_{\rm S}= \sqrt{m_{{\tilde t}_L}m_{{\tilde t}_R}}$.
Note that one loop radiative corrections to the CP-even Higgs mass depend logarithmically on  the stop quark mass and linearly on $X_t$.

 From the $A_t - m_{\tilde{t}_R}$  plane    in Figure \ref{atmstop},   we see that the in the mGMSB model,  $A_t/m_{\tilde{t}_R}<1$ is always the case,  no matter how large the values of    $M_{\mathrm{mess}}$ and  $\Lambda$.
 As described in section \ref{model}, the A-terms in this model vanish at the messenger scale whereas the scalar masses are given by Eq. (\ref{scalarmass}).
  The RGE running, however,  can generate large values for the A-terms at low scale. But as we see from Eq. (\ref{mh})  and (\ref{A1}), for radiative corrections to the Higgs mass the ratio, $A_t/m_{\tilde{t}_R}$ is important and not the actual value of $A_t$.
 As shown explicitly in ref.\cite{Gogoladze:2009bd},
 with $A_t/m_{\tilde{t}_R}<1$, the  radiative corrections to the lightest CP-even Higgs mass  are dominantly  generated   from   logarithmic corrections.  This explains why the stop quark masses have to lie in the few TeV region  in the mGMSB model. We also observe that $m_{\tilde{t}_R}$ becomes lighter for the $n_5=5$ case.  We can   understand this from Eq. (\ref{gauginomass}) which shows that the MSSM gaugino masses increase by a factor 5 if we increase $n_5$ from 1 to 5. According to the Eq. (\ref{scalarmass}), however, the scalar masses only increase by a factor $\sqrt{n_5}=\sqrt{5}$.  This  means that the large gaugino, particularly the gluino, mass enhances the low scale value of  $A_t$ through  RGE running. This explains why for $n_5=5$, we have more red points around the unit-slope line in the $A_t - m_{\tilde{t}_R}$ plane, which, on the other hand, also relaxes the lower bound on the  stop quark mass.

\begin{figure}
\begin{center}

\includegraphics[width=7.2cm,height=6.cm]{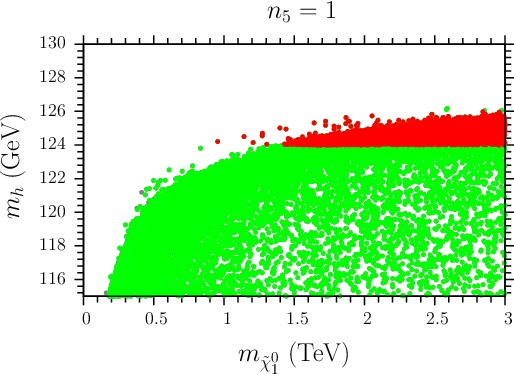}
\includegraphics[width=7.2cm,height=6. cm]{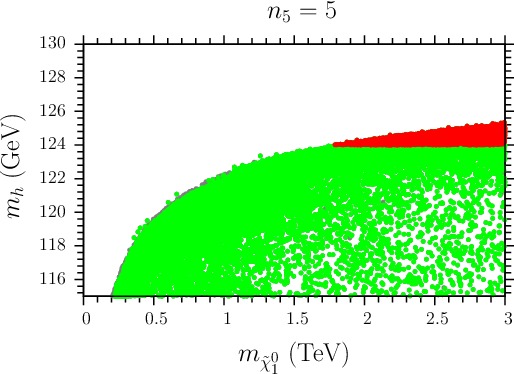}\vspace*{3mm}
\includegraphics[width=7.2cm,height=5.6cm]{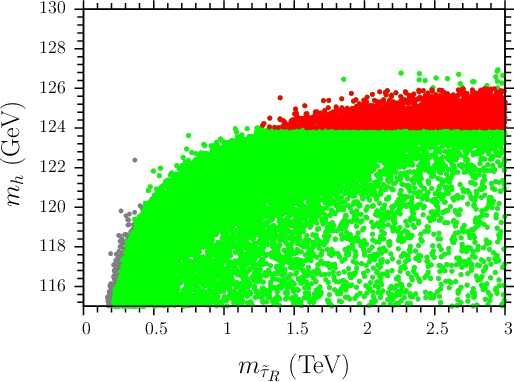}
\includegraphics[width=7.2cm,height=5.6cm]{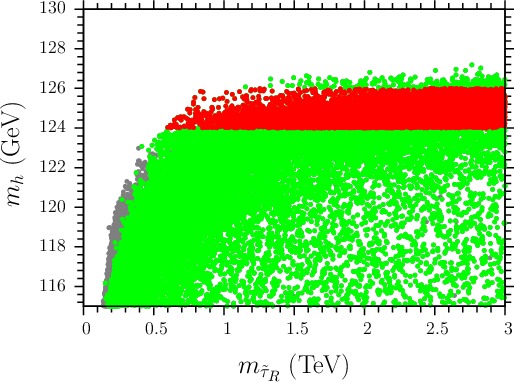}\vspace*{3mm}
\includegraphics[width=7.2cm,height=5.6cm]{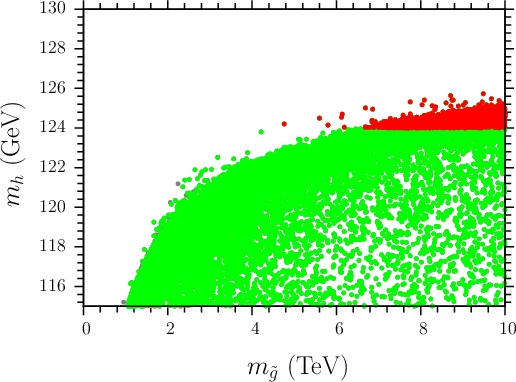}
\includegraphics[width=7.2cm,height=5.6cm]{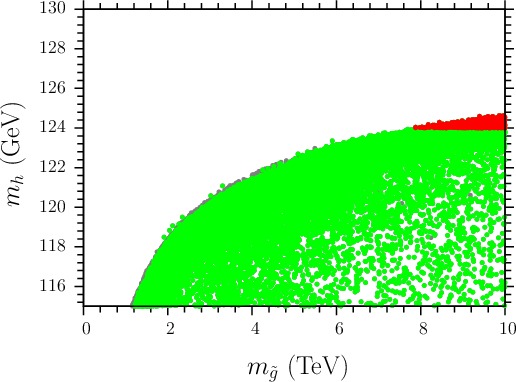}
\end{center}
\caption{Plots in $m_h - m_{\tilde{\chi}_1^0}$, $ m_h - m_{\tilde{\tau}_R} $ and $ m_h - m_{\tilde{g}} $ planes for $n_5=1$ and $n_5=5$. Color coding is the same as described in Figure \ref{fundp}.}
\label{fig3}
\end{figure}

 In Figure \ref{fig3} we  display our results in the $ m_h - m_{\tilde{\chi}_1^0} $, $m_h - m_{\tilde{t}_R}$ and $ m_h - m_{\tilde{g}} $ planes for $n_5=1$ and $n_5=5$.   Here $m_{\tilde{\chi}_1^0} $ and $m_{\tilde{g}} $ denote the lightest neutralino and gluino masses, respectively. The color coding is the same as in Figure \ref{fundp}. As mentioned earlier, Eqns. (\ref{gauginomass}) and (\ref{scalarmass}) show that the scalar masses scale as $\sqrt{n_5}$,  whereas the gaugino masses scale as $ n_5$, which is why the scalars are typically lighter than the gauginos for larger $n_5$ values.  This  explains why the lower bound  on
$m_{\tilde{\chi}_1^0} $ and $m_{\tilde{g}} $ increases for higher values  of $n_5$. Also, the lowering of the bound on $m_{\tilde{t}_R}$  is strongly related to how the stop mass changes, as discussed above in analyzing Figure  \ref{atmstop}. The lightest gluino for $n_5=1$ is $\sim$ 4.5 TeV, whereas for $n_5=5$, the lower bound on $m_{\tilde{g}}$ increases up to $\sim$ 8 TeV.

 As all other sparticles, the lightest MSSM neutralino is also heavy with a minimum mass  $\sim 1$ TeV for $n_5=1$, and $\sim 1.8$ TeV for $n_5=5$.
  For $n_5=1$, the neutralino is typically the NLSP in mGMSB. Since all other sparticles are much heavier, the neutralino, which is essentially a bino, dominantly decays to a gravitino and  photon ($\tilde{G}\gamma$). Other decay channels $\tilde{\chi}^0_1 \rightarrow \tilde{G} Z$ and $\tilde{\chi}^0_1 \rightarrow \tilde{G} h$ are also open but relatively suppressed.

The  collider signals for mGMSB at the Tevatron and LHC were studied in \cite{Feng:2010ij, Baer:1996hx}. Neutralino pair production  at the LHC can take place via loop suppressed gluon fusion or through subprocesses $q \overline{q} \rightarrow \tilde{\chi}^0_i \tilde{\chi}^0_j$ \cite{Gounaris:2004fm}, which yield tiny cross sections for large neutralino and squark masses. Single neutralino production in association with a squark, gluino or a chargino is also suppressed. If produced, the neutralino would lead to final states with photons plus missing energy, where the missing energy results from the gravitino.
 In reference  \cite{Feng:2010ij}, the expected number of events for prompt (non-prompt) di-photon (photon) events were estimated for the NLSP neutralino. It was shown that $N_{\gamma \gamma}(N_{\gamma})\lesssim 1$ for neutralino mass $\sim 1 \rm \ TeV$ for 14 TeV LHC with $10 {\rm \  fb^{-1}}$ integrated luminosity. A search for mGMSB model in final states with diphoton events and missing transverse energy  was performed by the CDF \cite{Aaltonen:2009tp} and D0 \cite{Abazov:2010us} collaborations and no excess above the SM expectations was observed. The limits on the sparticle masses obtained from the Higgs mass bound that we have found are far more stringent compared to those obtained from these searches.

Comparing the $ m_h - m_{\tilde{\chi}_1^0} $ and  $m_h - m_{\tilde{\tau}_R}$ planes in Figure \ref{fig3}, we can observe that
for $n_5=5$, $\tilde{\tau}_R$ can be the NLSP, with a minimum value  $\sim 600$ GeV. The NLSP stau dominantly decays to $\tilde{G} \tau$.
 It was shown in reference \cite{Feng:2010ij} that the event yield for $m_{\tilde{\tau}} > 600 \rm \ GeV$ is less than 10 for 14 TeV LHC with $10 {\rm \  fb^{-1}}$ integrated luminosity, for final states with non-prompt and metastable leptons. Thus, it will be very difficult to see any events characteristic of the mGMSB scenario at the LHC if the current preferred value of the Higgs mass ($\sim 124 -126 \rm \ GeV$) is confirmed.
The pseudo-scalar Higgs boson of MSSM turns out to have a mass $\gtrsim 3 $ TeV. The large limit on $m_A$ also implies that the lightest CP-even Higgs $h$ is very much SM-like.

The gravitino, which is the spin 3/2 superpartner of the graviton, acquires mass through spontaneous  breaking of local supersymmetry. The gravitino mass in such scenarios can be $\sim 1 {\rm \ eV} - 100 {\rm \ TeV}$. A light gravitino is a plausible dark matter candidate and can also manifest itself through missing energy in colliders \cite{Feng:2010ij}. In mGMSB the gravitino mass is given by
\begin{eqnarray}
m_{\tilde G}={F \over \sqrt{3} M_P}=2.4\left(\sqrt{F} \over 100 {\rm \ TeV}\right)^2 {\rm eV},
\label{gravitino_mass}
\end{eqnarray}
where the reduced Planck scale $M_P=2.4 \times 10^{18} \rm \ GeV$. The lower limit on $\Lambda$ and $M_{\rm mess}$ implies a lower limit on the gravitino mass.

\begin{figure}
\begin{center}

\includegraphics[width=7.2cm,height=5.6cm]{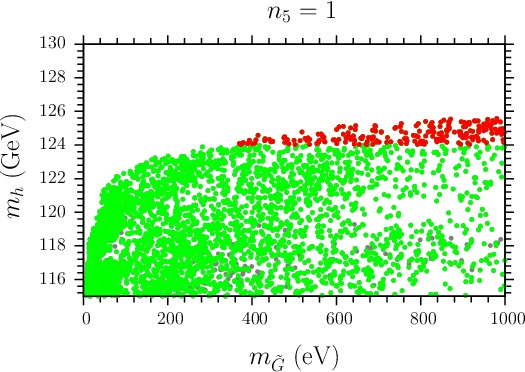}
\includegraphics[width=7.2cm,height=5.6cm]{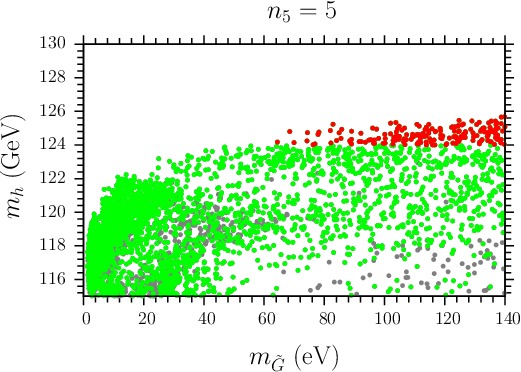}
\end{center}
\caption{Plots in $m_h - m_{\tilde G}$,  planes for $n_5=1$ and $n_5=5$. Color coding is the same as described in Figure \ref{fundp}.}
\label{fig4}
\end{figure}

We present in Figure \ref{fig4}  the results in $m_h - m_{\tilde G}$  planes for $n_5=1$ and $n_5=5$. The color coding is the same as described in Figure \ref{fundp}.
For $n_5=1$, the lower limit ($\sim 10^6 \rm \ GeV$) on these parameters  implies that the lightest allowed gravitino mass $\sim$ 360 eV. The Higgs mass window in Eq. (\ref{mhrange}) therefore excludes very light gravitinos which can be produced in the standard cosmological scenarios.
In standard scenarios, the relic density bound ($\Omega h^2 \sim 0.11$ \cite{Komatsu:2008hk}) is satisfied with a gravitino mass $\sim$ 200 eV \cite{Feng:2010ij}, which makes it a hot dark matter candidate.
 The hot component of dark matter, however, cannot be more than $15\%$ which in turn implies that the gravitino mass $\lesssim 30$ eV    \cite{Feng:2010ij}. In standard  scenarios, therefore, the gravitino can form only a fraction of dark matter. A gravitino mass $\gtrsim$ 30 eV requires non-standard scenarios  in order to agree with observations.
 Such non-standard scenarios include gravitino decoupling and freezing out earlier than in the standard scenario, which may be possible in a theory with more degrees of freedom than the MSSM \cite{Feng:2010ij}. For $n_5=1$, the lightest gravitino can be $\sim 360$ eV. For $n_5=5$, however, the lower limits on $\Lambda$ and $M_{\rm mess}$ are smaller and the gravitino mass can be as light as $\sim 60$ eV.  A gravitino of mass $\gtrsim {\rm \ keV}$ is still possible for $n_5=1 \text{ or }5$, and it can be cold enough to constitute all of the dark matter if non-standard scenarios such as early decoupling is assumed. Note that these lower bounds on the gravitino mass apply for $c_{\rm grav}=1$. For $c_{\rm grav}>1$, the lower bound on $m_{\tilde{G}}$ will increase, which  will make the gravitino problem more severe.

In Table \ref{tab1}, we show three benchmark points satisfying the various constraints mentioned in section \ref{constraintsSection}. These display  the minimal values of the neutralino, stau and gravitino masses in mGMSB that are compatible with a 125 GeV CP-even Higgs boson. Point 1 shows that the lightest NLSP neutralino allowed mass is  around 1.4 TeV for $n_5=1$. The second point has the lightest stau that can be realized for $n_5=5$. The last point shows a 686 eV gravitino, which is the lightest value we found for a Higgs mass $\sim 125 \rm \ GeV$. The rest of the spectrum turns out to be quite heavy, as expected, for all  three benchmark points, with the squarks typically heavier than 10 TeV and the sleptons have masses more than 2 TeV.

\begin{table}[t!]\vspace{1.5cm}
\centering
\begin{tabular}{|p{3cm}|p{3cm}p{3cm}p{3cm}|}
\hline
\hline
                 	&	 Point 1 	&	 Point 2 	&	 Point 3 	\\
\hline

$\Lambda $         	&$	     1\times 10^{6}	$&$	          4.22\times 10^{5}	$&$	           1.5\times 10^{6}	$\\
$M_{\rm mess} $         	&$	     1.74\times 10^{14}	$&$	          9.02\times 10^{12}	$&$	           1.9\times 10^{6}	$\\
$n_5$         	&$	1	$&$	5	$&$	1	$\\
$\tan\beta$      	&$	42	$&$	60	$&$	46	$\\
\hline		  		  		  	
$\mu$            	&$	7873	$&$	5802	$&$	3678	$\\

$m_h$            	&$	\textbf{125}	$&$	\textbf{125.2}	$&$	\textbf{125.1}	$\\
$m_H$            	&$	9930	$&$	4865	$&$	5141	$\\
$m_A$            	&$	9865	$&$	4833	$&$	5107	$\\
$m_{H^{\pm}}$    	&$	9930	$&$	4866	$&$	5142	$\\
		  		  		  	
\hline		  		  		  	
$m_{\tilde{\chi}^0_{1,2}}$	&$	        \mathbf{1398},         2619	$&$	        2924,         5307	$&$	        2405,         3732	$\\

$m_{\tilde{\chi}^0_{3,4}}$	&$	        7775,         7775	$&$	        5833,         5836	$&$	        3735,         4449	$\\

$m_{\tilde{\chi}^{\pm}_{1,2}}$	&$	        2624,         7711	$&$	        5315,         5840	$&$	        3811,         4364	$\\

$m_{\tilde{g}}$  	&$	6689	$&$	12312	$&$	10613	$\\
		  		  		  	
\hline $m_{ \tilde{u}_{L,R}}$	&$	       15956,        14113	$&$	       12014,        11276	$&$	       14064,        13301	$\\
                 		  		  		  	
$m_{\tilde{t}_{L,R}}$	&$	       13637,         9847	$&$	       10289,         8994	$&$	       13027,        11873	$\\
                 		  		  		  	
\hline $m_{ \tilde{d}_{L,R}}$	&$	       15956,        13540	$&$	       12015,        11151	$&$	       14064,        13213	$\\
                 		  		  		  	
$m_{\tilde{b}_{R}}$	&$	12233	$&$	9720	$&$	12421	$\\
                 		  		  		  	
\hline		  		  		  	
$m_{\tilde{\nu}_{1}}$	&$	9281	$&$	4885	$&$	5112	$\\
                 		  		  		  	
$m_{\tilde{\nu}_{3}}$	&$	8722	$&$	4424	$&$	5009	$\\
                 		  		  		  	
\hline		  		  		  	
$m_{ \tilde{e}_{L,R}}$	&$	        9290,         6774	$&$	        4900,         2962	$&$	        5133,         2640	$\\
                		  		  		  	
$m_{\tilde{\tau}_{L,R}}$	&$	        8706,         5109	$&$	        4407,          \textbf{783}	$&$	        4991,         2306	$\\

\hline							
							
$m_{\tilde{G}}$     	&$	42	$&$	0.916	$&$	\mathbf{6.86\times 10^{-7}}	$\\

\hline
\hline
\end{tabular}
\caption{ Benchmark points for the mGMSB. All masses are in units of GeV. Point 1, 2 and 3 show the lightest neutralino, stau and gravitino (shown in bold) that can be realised in mGMSB for a Higgs mass of 125 GeV. For the three points, $m_t=173.3 {\rm \ GeV}$ and $c_{\rm grav}=1$. }
\label{tab1}
\end{table}


\section{Conclusion \label{conclusions}}

The ATLAS and CMS experiments at the LHC have presented tantalizing albeit tentative evidence for the existence of the SM Higgs boson with mass close to 125 GeV. We have explored in this paper the implications of this observation for the sparticle spectroscopy of the minimal gauge mediated supersymmetry breaking scenario. By performing a random scan of the fundamental parameter space, we find that accommodating a 125 GeV Higgs mass in these models typically forces the sparticle spectrum, with few exceptions, to lie in the few to multi-TeV mass range. The colored sparticles, in particular, all have masses in the multi-TeV range.

With a single $5+\overline{5}$   pair of SU(5) messenger fields, the lightest MSSM neutralino mass lies close to 1 TeV. As we increase the number of SU(5) messenger multiplets the MSSM gauginos, and hence the neutralino, become heavier.
  The lightest stau mass is close to 1.4 TeV  for the single $5+\overline{5}$  models, and it becomes lighter  as we increase the number of SU(5) messenger multiplets. Particularly, with five pairs of $5+\overline{5}$, stau becomes the NLSP and can be as light as 800 GeV.
 The detection of a stau at the LHC may shed light on the number of SU(5) messenger multiplets at the messenger scale.

A Higgs mass close to 125 GeV also yields lower limits on both the messenger and soft supersymmetry breaking scales which, in turn, constrain the gravitino mass. A single $5+\overline{5}$  pair requires that the gravitino mass  $\gtrsim 360$ eV. With five pairs of  $5+\overline{5}$  messenger fields, this lower limit on the gravitino mass is reduced to 60 eV. The simplest GMSB models, it appears, require non-standard cosmological scenarios in order to be in agreement with observations \cite{Feng:2010ij}.

\section*{Acknowledgments}
We thank  Azar Mustafayev   for valuable discussions and comments.
This work is supported in part by the DOE Grant No. DE-FG02-91ER40626. This work used the Extreme Science
and Engineering Discovery Environment (XSEDE), which is supported by the National Science
Foundation grant number OCI-1053575.



\end{document}